%% file: main2.tex
\documentclass[conference]{IEEEtran}

\usepackage{cite}
\usepackage{amsmath,amssymb,amsfonts}
\usepackage{algorithmic}
\usepackage{graphicx}
\usepackage{textcomp}
\usepackage{xcolor}
\usepackage{url}
\usepackage{booktabs}
\usepackage{tabularx}
\usepackage{multirow}
\usepackage{array}
\usepackage{microtype}
\usepackage{balance}

\usepackage{mdframed}
\usepackage{courier} 
\usepackage[utf8]{inputenc}
\usepackage{tikz}
\usetikzlibrary{positioning, fit, calc, matrix}

\def\BibTeX{{\rm B\kern-.05em{\sc i\kern-.025em b}\kern-.08em T\kern-.1667em\lower.7ex\hbox{E}\kern-.125emX}}

\def\BibTeX{{\rm B\kern-.05em{\sc i\kern-.025em b}\kern-.08em
    T\kern-.1667em\lower.7ex\hbox{E}\kern-.125emX}}

\begin{document}

\title{RF-Analyzer: Can Vision-Language Models Learn RF Understanding from Synthetic Data?}

\author{
\IEEEauthorblockN{Anis Bara, Lina Bariah, Hang Zou,  Brahim Mefgouda, Merouane Debbah}
\IEEEauthorblockA{
\textit{Research Institute for Digital Future, Khalifa University} \\
anis.bara@ku.ac.ae, lina.bariah@ku.ac.ae, hang.zou@ku.ac.ae, brahim.mefgouda@ku.ac.ae,   merouane.debbah@ku.ac.ae
}
}

\maketitle
\begin{abstract}
Understanding the wireless spectrum is a fundamental requirement for intelligent communication systems, however, interpreting spectrograms requires extracting multiple physical attributes and reasoning about signal structure, which is a capability that is not achieved by traditional ML approaches. Recent advances in vision-language models (VLMs) demonstrated the possibility of learning such interpretation capabilities directly from data. This paper investigates whether VLMs can learn this capability from synthetic data alone, and more importantly, whether such learned representations generalize to real over-the-air RF environments. To address this question, we introduce RF-Analyzer, an SDR-to-AI analysis platform that integrates live spectrum captures associated with the corresponding VLM-based interpretation, enabling direct evaluation of VLMs outputs on live over-the-air signals. Using this platform, we assess a model trained exclusively on synthetic spectrogram data with general-purpose baselines. To enable systematic analysis, we establish a benchmark framework comprising three metrics, Physical Attribute Extraction Score (PAES), Prompt Leakage Rate (PLR), and hallucination count, to assess signal understanding and grounding. The obtained results demonstrate that VLMs trained on synthetic spectrogram data can generalize to real RF environments, particularly for extracting physical signal attributes such as spectral occupancy, temporal behavior, and SNR. This indicates that synthetic data is sufficient for learning transferable representations of RF signal structure. However, this generalization is limited due to the fact that synthetic training does not provide reliable semantic grounding without contextual priors. In particular, generalization breaks under conditions that are not covered in the synthetic distribution, particularly low-SNR regimes. 
\end{abstract}


\section{Introduction}

Large language models (LLMs) have demonstrated strong performance across a broad range of reasoning and planning tasks, enabled by scale, instruction tuning, and retrieval augmented generation in telecom domains \cite{bariah2024largegenerativeai,zou2025telecomgpt,zhou2025llmtelecom}. Despite these advances, LLMs remain fundamentally limited by their reliance on tokenized representations and lack the direct grounding in physical environments. This limitation is particularly critical in wireless networks, where decision-making depends not only on protocol-level knowledge but also on the instantaneous electromagnetic state of the radio spectrum. In particular, key features, such as signal presence, spectral occupancy, temporal behavior, and signal-to-noise ratio (SNR) levels, must be captured from raw RF observations rather than textual inputs. As a result, these signals must first be sensed and transformed into structured representations before they can be used by higher-level reasoning systems, motivating the need for physically grounded AI models for spectrum understanding.

Before network-level decisions can be made, the electromagnetic environment must be observed, characterized, and represented in a form that higher-level models can process. Traditionally, this sensing function is realized through two main categories of tools. The first consists of dedicated hardware instrumentation, such as spectrum analyzers and signal monitoring receivers~\cite{rohdeschwarz2024, keysight2024}, which provide calibrated, high-dynamic-range measurements. However, these systems rely heavily on manual configuration and expert interpretation, producing outputs that are not readily structured or machine-readable for integration into automated pipelines. The second category includes task-specific machine learning approaches, where convolutional and recurrent architectures, trained on labeled IQ datasets~\cite{oshea2018over, boegner2022largescalerf}, automate modulation classification. While these models achieve high accuracy within their training distributions, they are limited to predicting discrete labels from predefined label spaces, offering limited interpretability, no natural-language interface, and limited robustness to variations in SNR and channel conditions, that are not seen in the training datasets~\cite{swami2000hierarchical, west2017deep}. As a result, both paradigms are not sufficient to enable flexible and structured understanding of RF signals. Interpreting spectrograms requires extracting multiple physical attributes and reasoning about signal structure, rather than assigning a single class label. Recent advances in vision-language models (VLMs) have the potential to bridge this gap, as they can jointly process visual inputs and generate structured natural-language descriptions. However, their ability to operate on domain-specific representations such as RF spectrograms is still in its early stages.

Recent research attempts were developed to address this gap by adapting VLMs to the RF domain. Earlier studies have shown that transforming IQ signals into spectrogram-based representations and applying parameter-efficient fine-tuning enables these models to capture RF signal patterns effectively~\cite{zou2025seeingradiokhalifa}. Extensions such as RF-GPT~\cite{zou2026rf} further demonstrate that VLMs can support a broader range of RF-related tasks, including wireless technology identification, signal analysis, and information extraction. These efforts demonstrate that VLMs can learn structured RF representations through targeted fine-tuning, particularly when combining spectral and temporal information. However, despite these advancements, the generalization of such representations beyond synthetic training data is not yet properly understood. In particular, it is unclear to what extent models trained exclusively on synthetic RF data can provide reliable performance when exposed to real over-the-air environments, where signal conditions, noise characteristics, and propagation effects differ significantly from simulated settings. It is worthy to highlight that real RF data are expensive to collect, difficult to label, location-dependent, and usually constrained by spectrum regulations. If synthetic-only training can support real spectrum understanding, then RF language models can be developed and further tuned more efficiently, compared to systems that rely on large-scale labeled over-the-air datasets.

Synthetic datasets offer accessibility, scalability, controllability, and precise labeling, rendering them highly attractive for training large models. However, real-world RF environments exhibit unique characteristics that are difficult to fully replicate in simulation, including heterogeneous interference patterns, hardware-based impairments, and dynamic channel conditions. As a result, the ability of VLMs to transfer learned representations from synthetic to real domains becomes a key factor in determining their applicability to real-world spectrum analysis tasks.


To study this gap, in this paper, we present \textbf{RF-Analyzer}, a GPU-accelerated spectrum analysis platform that connects an Ettus USRP~B210 software-defined radio to multiple VLM backends, including domain-adapted and general-purpose models, such as RF-GPT, Qwen2.5-VL, and Llama~3.2~Vision. RF-GPT is a VLM fine-tuned on synthetic spectrogram data \cite{zou2026rf}. Unlike conventional spectrum analyzers, RF-Analyzer does not operate around predefined label spaces or fixed measurement procedures, but rather it functions as a reasoning engine, accepting a rendered waterfall display as input and producing structured natural-language descriptions of the signal’s physical attributes, including modulation family, spectral occupancy, temporal duty cycle, SNR class, and channel isolation, through a conversational interface.

The main contributions of this paper are as follows:
\begin{itemize}
    \item We present \textbf{RF-Analyzer}, an integrated SDR-to-AI platform connecting a USRP~B210 receiver to a locally hosted VLM backend, enabling evaluation of RF spectrogram interpretation on real over-the-air signals. 
    \item We provide a synthetic-to-real evaluation of RF-trained VLMs (such as RF-GPT), assessing whether models trained only on synthetic spectrogram data can generalize to live RF environments without any real-data fine-tuning.
    \item We propose the \textbf{Physical Attribute Extraction Score (PAES)}, which is a structured evaluation framework assessing model outputs across five signal description dimensions, quantifying RF signal understanding beyond classification accuracy.
    \item We introduce \textbf{Prompt Leakage Rate (PLR)} and hallucination counting to assess to what extent model outputs are grounded in the input spectrogram, rather than influenced by prompt metadata or unsupported claims.
    \item We characterize training-set label bias, as a reproducible failure mode in domain-adapted VLMs, demonstrating that a model can correctly perceive signal geometry while misclassifying a technology label due to incomplete frequency-to-standard coverage in the fine-tuning data.
\end{itemize}

The remainder of this paper is organized as follows. Section~\ref{sec:rfanalyzer} details the RF-Analyzer platform.  Section~\ref{sec:exp} presents the experimental evaluation, including setup, metrics, results, and the paper is concluded in Section~\ref{sec:conclusion}.

\section{RF-Analyzer: The Evaluation Platform}
\label{sec:rfanalyzer}


RF-Analyzer is a real-time spectrum analysis platform designed to bridge the gap between software-defined radio measurements and VLM-based RF reasoning. In this work, RF-Analyzer acts as the evaluation front-end for assessing whether a model trained on synthetic RF spectrograms can interpret live over-the-air captures, while enabling direct comparison with general-purpose VLMs under identical input conditions. The design principle ensures that model predictions are derived from the same waterfall display and instrument metadata available to a human spectrum analyst, rather than from hidden labels or precomputed classifier outputs.

\subsection{RF-GPT Backbone}

RF-Analyzer uses RF-GPT as its primary domain-adapted VLM backend. RF-GPT is a radio-frequency language model built on top of Qwen2.5-VL and adapted to interpret RF spectrograms rather than natural images. Complex IQ waveforms are first converted into time-frequency representations, which are then processed by the vision encoder and passed to the language model as RF-grounded visual tokens. The model is trained to answer natural-language questions about RF scenes, including signal type, occupied bandwidth, temporal behavior, SNR regime, overlap, and wireless-technology attributes~\cite{zou2026rf}.

It is important to emphasize that RF-GPT is used \textit{without any real-data tuning}, where its primary fine-tuning corpus is fully synthetic. The considered standards-compliant waveform generators produce wideband RF scenes, spectrograms, configuration metadata, and relevant captions, which are then converted into instruction–answer examples. Hence, there are no live over-the-air captures from the experiments in Section~\ref{sec:exp} are used for additional fine-tuning, calibration, or few-shot
adaptation. Thus, the evaluation in this paper specifically measures synthetic-to-real generalizability, focusing on whether RF structure learned from simulated spectrograms adapts to waterfall images captured by a real SDR front-end.

\subsection{Signal Acquisition}


The application interfaces with an Ettus USRP~B210 via the UHD~4.x driver. The B210 covers 70\,MHz to 6\,GHz with up to 56\,MHz of instantaneous bandwidth and a 12-bit ADC. In hardware mode, complex baseband IQ samples are streamed continuously from the receiver into the RF-Analyzer processing pipeline. When no hardware is present, the application can instead use a software signal simulator that injects FM-like, LTE-like, and pulsed waveforms for interface testing and software development. This simulator is not used for the over-the-air evaluation reported in Section~\ref{sec:exp}.

\begin{figure}[t!]
  \centering
  \resizebox{\columnwidth}{!}{%
    \input{figures/architecture}%
  }
  \caption{System architecture of RF-Analyzer. Colored blocks indicate
  functional layers: hardware (orange), acquisition and DSP (teal),
  display (blue), AI engine (violet), and application shell (gray).
  Solid arrows show streaming data flow; the dashed arrow denotes the
  user-triggered analysis path from \texttt{MainWindow} to
  \texttt{AIClassifierManager}.}
  \label{fig:architecture}
\end{figure}
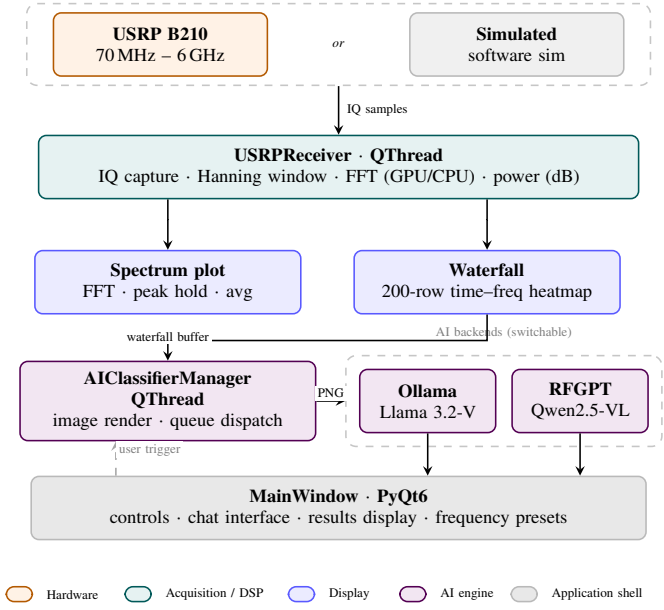

\subsection{DSP Pipeline}


The received complex baseband samples are processed in blocks of 2048 points. The pipeline applies a Hann window, computes the FFT, optionally accelerates the operation on the GPU via CuPy/CUDA, converts the magnitude spectrum to dB scale, and averages multiple frames to reduce display noise. The processed spectra are displayed on a live spectrum plot and appended to a rolling waterfall buffer with 200 time rows. A threshold-based peak detector is used for operator visualization, but its detected labels are not passed to the VLM. The AI backend receives only the rendered waterfall image and the measurement metadata explicitly included in the prompt.





\subsection{AI Backend Architecture}

The RF-Analyzer supports multiple VLM backends through a unified interface so that domain-adapted and general-purpose VLMs can be evaluated under identical input conditions. Each backend receives the same rendered waterfall PNG image and the same text prompt, ensuring that none of the VLMs under consideration have access to raw IQ samples, peak-detector outputs, ground-truth labels, or scenario identifiers.

Two categories of backends are considered in this work. The first corresponds to general-purpose VLMs, which process the spectrogram image through a vision-capable interface served locally via Ollama- or vLLM-compatible frameworks, without domain-specific adaptation. In this work, this backend is used to evaluate models such as Llama~3.2~Vision~11B and the untuned Qwen2.5-VL baseline. The second corresponds to a domain-adapted VLM trained on synthetic RF spectrogram data, loaded via vLLM~\cite{kwon2023efficient} and executed on a local GPU, with end-to-end analysis latency on the order of seconds per query. This backend generates natural-language descriptions of the observed RF scene based on the input spectrogram. 



To interface with the AI backend, RF-Analyzer renders the current waterfall buffer as a Matplotlib figure with calibrated frequency axis, color-mapped power scale, and a title containing the center frequency and sample rate. The resulting PNG image, together with the text prompt, is then dispatched to the selected backend. The full GUI screenshot, previous model responses, and red figure annotations shown in Fig.~\ref{fig:rfanalyzer} are not included in the model input.

\begin{figure}[t]
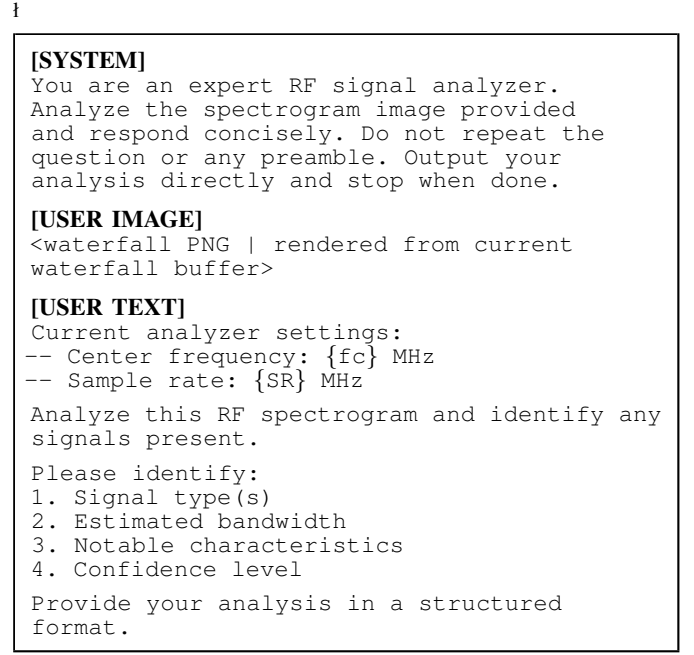

\footnotesize

\l
\begin{mdframed}[linewidth=0.8pt, innerleftmargin=6pt, 
                 innerrightmargin=6pt, innertopmargin=6pt, 
                 innerbottommargin=6pt]

{\small\textbf{[SYSTEM]}} \\
{\small\ttfamily You are an expert RF signal analyzer. Analyze 
the spectrogram image provided and respond concisely. Do not 
repeat the question or any preamble. Output your analysis 
directly and stop when done.}

\medskip
{\small\textbf{[USER IMAGE]}} \\
{\small\ttfamily <waterfall PNG — rendered from current 
waterfall buffer>}

\medskip
{\small\textbf{[USER TEXT]}} \\
{\small\ttfamily Current analyzer settings:} \\
{\small\ttfamily \quad -- Center frequency: \{fc\} MHz} \\
{\small\ttfamily \quad -- Sample rate: \{SR\} MHz} \\[4pt]
{\small\ttfamily Analyze this RF spectrogram and identify any 
signals present.}\\[4pt]
{\small\ttfamily Please identify:}\\
{\small\ttfamily \quad 1.\ Signal type(s)}\\
{\small\ttfamily \quad 2.\ Estimated bandwidth}\\
{\small\ttfamily \quad 3.\ Notable characteristics}\\
{\small\ttfamily \quad 4.\ Confidence level}\\[4pt]
{\small\ttfamily Provide your analysis in a structured format.}

\end{mdframed}
\caption{Prompt template sent to the AI backend on each 
analysis request. The system message is injected automatically; 
\{fc\} and \{SR\} are substituted with the current hardware 
settings at runtime. The center frequency and sample rate are 
the only numeric values accessible to the model before image 
inspection, and are the source of the prompt leakage measured 
by PLR.}
\label{fig:prompt}
\end{figure}



\subsection{User Interface}

\begin{figure}[h]
    \centering
    \includegraphics[width=0.94\columnwidth]{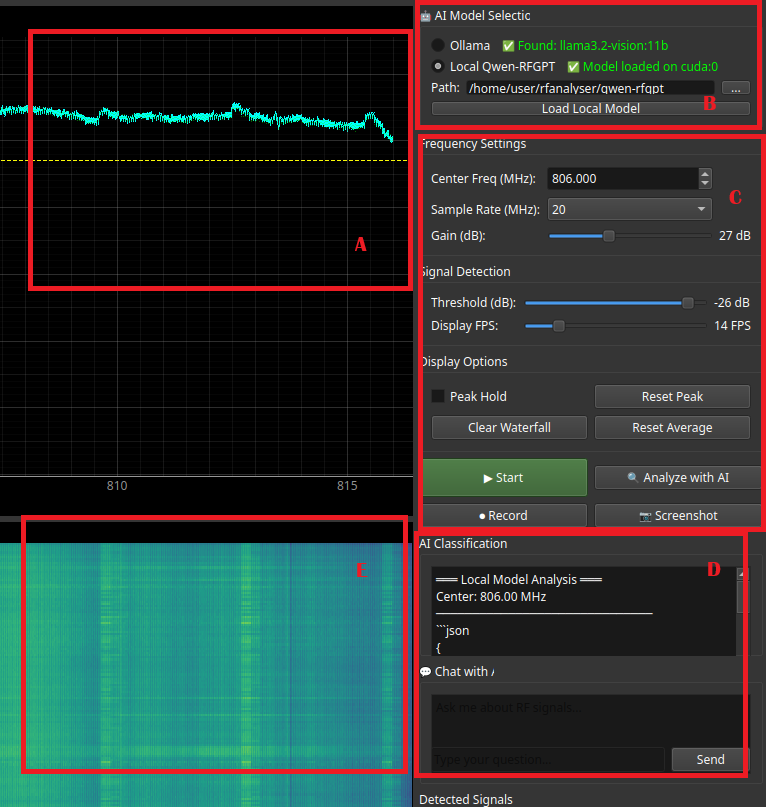}
    \caption{RF~Analyzer running on an Ubuntu~24 workstation with an Ettus USRP~B210. The top panel shows the live spectrum at 806\,MHz (A) with a 20\,MHz sample rate (C). The waterfall (bottom panel E) reveals a broadband continuous signal (the bright block) and a wideband interference event (the vertical purple stripe). The AI analysis panel (right) shows RF-GPT's assessment of the capture (D).}
    \label{fig:rfanalyzer}
\end{figure}

The user interface, shown in Fig.~\ref{fig:rfanalyzer}, consists of a spectrum plot at the top (\textbf{A}), a waterfall display below (\textbf{E}), and a control panel on the right (\textbf{B} and \textbf{C}). The control panel exposes center frequency, sample rate, gain, detection threshold, FFT size, and
display FPS, as well as quick-access presets for common bands from 98\,MHz FM broadcast to 2.4\,GHz Wi-Fi. The AI analysis panel and free-form chat window appear below the controls (\textbf{D}), allowing the users to ask follow-up questions without leaving the application. The screenshot in Fig.~\ref{fig:rfanalyzer} illustrates the complete operator interface; the VLM input is restricted to the separately rendered waterfall image described above.

\section{Experimental Evaluation}
\label{sec:exp}

\subsection{Setup}

All experiments were conducted in Abu Dhabi, UAE, using the USRP B210 connected to a broadband log-periodic antenna. Five test scenarios were designed to cover the training distribution spectrum of RF-GPT, from fully in-distribution signals (the signal captured in the real world belongs to the same category of signals that RF-GPT saw during synthetic training, while out-of-distribution means unseen signals) to signals at the boundary of the synthetic corpus. A sixth adversarial scenario was added as an out-of-distribution (OOD) stress test. Table~\ref{tab:scenarios} summarizes the hardware settings for each scenario.

We use the term \emph{category-aligned} to indicate that the real-world capture belongs to a broad signal family represented in the synthetic training corpus. This does not mean that the model was trained on the same location, transmitter, receiver front-end, antenna, or propagation environment. Accordingly, the evaluation focuses on assessing whether RF-GPT transfers from synthetic spectrograms to real SDR waterfall captures without any real-data fine-tuning.

In the \textit{killer test} (KT), we placed a single Bluetooth 2.4\,GHz wireless mouse inside a commercially available RF-shielding bag. This arrangement produces a controlled scene containing exactly one transmitter operating at a power level attenuated by the bag whose spectral signature (a narrow, frequency-hopping burst centered near one of the 79 Bluetooth channels) is absent from the RF-GPT training corpus in its attenuated form. The test is designed to be adversarial. The prompt declares a center frequency of 2400\,MHz and a sample rate of 40\,MHz, providing misleading headroom for the bandwidth estimate.

\begin{table}[h]
\centering
\caption{USRP B210 Capture Settings per Scenario}
\label{tab:scenarios}

\begin{tabularx}{\columnwidth}{c r r r X}
\toprule
\textbf{ID} & \textbf{$f_c$} & \textbf{SR} & \textbf{Gain} & \textbf{Environment} \\
\midrule
S1 & 806\,MHz    & 20\,MHz & 30\,dB & LTE B20 downlink \\
S2 & 98\,MHz     & 10\,MHz & 25\,dB & FM broadcast \\
S3 & 433.92\,MHz & 5\,MHz  & 35\,dB & ISM pulsed devices \\
S4 & 2437\,MHz   & 40\,MHz & 20\,dB & Wi-Fi 2.4\,GHz ch.\,6 \\
S5 & 950\,MHz    & 20\,MHz & 30\,dB & GSM/LTE 900 \\
KT & 2400\,MHz   & 40\,MHz & 20\,dB & BT mouse in RF shield bag\\
\bottomrule
\end{tabularx}
\end{table}


Each scenario was captured three times without changing the hardware settings. All models received the same rendered waterfall image and the same text prompt, which listed only the center frequency and sample rate. The prompt did not reveal the signal type, bandwidth, modulation, SNR, or ground-truth scenario label.

\subsection{Evaluation Metrics}
 
To assess RF spectrogram interpretation, we define a set of complementary metrics that capture both structured signal understanding and grounding. These metrics are designed to evaluate whether model outputs reflect the physical properties of the observed RF scene and that are derived from the input spectrogram instead of from the prompt artifacts.
 
\subsubsection{Physical Attribute Extraction Score (PAES)}
Each response is evaluated against five binary attributes, derived from the ground-truth signal:
\begin{enumerate}
\item[$a_1 =$] \textit{Temporal class}: continuous / pulsed / burst, read from the waterfall time axis.
\item[$a_2 =$] \textit{Spectral occupancy class}: narrow ($<5$\,MHz) / medium (5--15\,MHz) / wide ($>15$\,MHz), measured from the spectrogram, not from the sample-rate metadata.
\item[$a_3 =$] \textit{SNR class}: low / medium / high, inferred from the color contrast between the signal and the noise floor.
\item[$a_4 =$] \textit{Signal isolation}: isolated (single emitter) / overlapping (co-channel interference).
\item[$a_5 =$] \textit{Technology family}: cellular / ISM / broadcast / radar.
\end{enumerate}


An attribute scores 1 if the model's response matches the ground truth and 0 otherwise. The PAES for a trial is defined as $\sum_{i=1}^{5} a_i,$ and is therefore reported on a 0--5 scale. The scenario-level PAES is the average over three repeated captures.
 
\subsubsection{Prompt Leakage Rate (PLR)}
PLR measures whether a bandwidth estimate is derived from prompt metadata rather than inferred from the input image. A response is flagged as leaked when it reports a bandwidth within $\pm$1\,MHz of the sample-rate value provided in the prompt and either explicitly references the analyzer settings or provides no evidence grounded in the image. PLR is computed as the fraction of leaked bandwidth estimates among all responses that include a bandwidth estimate.
 
\subsubsection{Hallucination Count}
We count the number of factual claims per scenario that contradict measurable signal properties. The two most common types encountered in this study are (a) reporting no signal present when a signal is clearly visible in the waterfall (false-negative hallucination) and (b) predicting a specific technology label that is inconsistent with the frequency band and spectral shape.
 
\subsection{Ground Truth}
 
Ground truth was determined by manual inspection of the waterfall displays and validating the observations against UAE spectrum band plans. The evaluated scenarios are as follows. S1 (806\,MHz, LTE Band 20 downlink-a continuous, wide-band, medium-SNR cellular signal), S2 (FM broadcasting at 98\,MHz-a continuous, medium-band, medium-SNR broadcast signal), S3 (433.92\,MHz ISM-pulsed, narrow-to-medium band emissions consistent with remote-control or sensor devices), S4 (2437\,MHz, Wi-Fi channel~6 burst, wide-band, low-to-medium SNR), S5 (950\,MHz, GSM/LTE Band~8 downlink-continuous, wide-band, medium-high SNR), and KT (2400\,MHz, single Bluetooth mouse in a shielded bag-short burst, narrow-band, low SNR, isolated).

 
\begin{table*}[t!]
\centering
\caption{Per-Scenario PAES Scores and Representative Model Responses}
\label{tab:responses}
\renewcommand{\arraystretch}{1.1}
\small
\begin{tabular}{p{1.8cm} p{2.6cm} c p{9.5cm}}
\toprule
\textbf{Scenario} & \textbf{Model} & \textbf{PAES} & \textbf{Representative Response} \\
\midrule
S1 -- 806\,MHz & RF-GPT        & 4/5 & \textit{``Continuous block spanning the full 20\,MHz channel; high SNR; cellular family; no co-channel interference.''} \\
               & Qwen Base    & 0/5 & \textit{``No discernible signals; spectrogram appears uniformly yellow indicating very low power.''} \\
               & Llama 3.2V   & 2/5 & \textit{``Likely a CW or sinusoidal signal; estimated bandwidth approximately 10\,MHz; confidence low.''} \\
\midrule
S2 -- 98\,MHz  & RF-GPT        & 5/5 & \textit{``DVB-S2 BPSK broadcast stream; 10\,MHz bandwidth; medium SNR; isolated.''} \\
               & Qwen Base    & 0/5 & \textit{``No discernible signals; uniformly yellow coloration throughout the spectrum.''} \\
               & Llama 3.2V   & 4/5 & \textit{``Single strong signal at 98.0\,MHz; estimated bandwidth 10\,MHz; high SNR.''} \\
\midrule
S3 -- 433\,MHz & RF-GPT        & 5/5 & \textit{``Bluetooth Classic GFSK; continuous; high SNR; 5\,MHz; isolated.''} \\
               & Qwen Base    & 4/5 & \textit{``Continuous signal, medium power, 5\,MHz span; no technology identified.''} \\
               & Llama 3.2V   & 2/5 & \textit{``Narrowband signal at 433.92\,MHz; no temporal classification provided.''} \\
\midrule
S4 -- 2437\,MHz & RF-GPT       & 5/5 & \textit{``OOK signal detected; wide bandwidth; short burst; ISM family; no overlap.''} \\
                & Qwen Base   & 4/5 & \textit{``Single wide-band carrier identified; no technology attribution.''} \\
                & Llama 3.2V  & 3/5 & \textit{``Single signal; wide spectral occupancy; high SNR; no technology label.''} \\
\midrule
S5 -- 950\,MHz & RF-GPT        & 5/5 & \textit{``5G NR uplink, QPSK; full 20\,MHz; continuous; medium SNR; isolated.''} \\
               & Qwen Base    & 0/5 & \textit{``No discernible signals detected across all three trials.''} \\
               & Llama 3.2V   & 4/5 & \textit{``Single narrowband signal; approximately 20\,MHz BW; high SNR.''} \\
\midrule
KT -- 2.4\,GHz & RF-GPT       & 2/5 & \textit{``Two signals detected (OOK + QPSK); overlapping; busy scene.''} \\
(OOD)           & Qwen Base   & 0/5 & \textit{``No signal detected.''} \\
                & Llama 3.2V  & 0/5 & \textit{``No signal detected.''} \\
\bottomrule
\end{tabular}
\end{table*}
 
 

\begin{table}[t!]
\centering
\caption{Prompt Leakage Rate (PLR) and Hallucination Count}
\label{tab:plr}
\renewcommand{\arraystretch}{1.2}
\begin{tabular}{lccc}
\toprule
\textbf{Metric} & \textbf{RF-GPT} & \textbf{Qwen Base} & \textbf{Llama 3.2V} \\
\midrule
PLR (\%) & \textbf{15} & 55 & 70 \\
False-negative halluc. & 2 & 13 & 8 \\
Tech.-label halluc. & 6 & 0 & 1 \\
Total halluc. count & \textbf{8} & 13 & 9 \\
\bottomrule
\end{tabular}
\end{table}
 

\subsection{Result Analysis and Failure Modes}
Table~\ref{tab:responses} presents the PAES scores for all scenarios. The obtained results show that RF-GPT is the only model that consistently extracts visible RF structure from the live waterfall captures. The model demonstrates its strongest performance in estimating spectral occupancy and identifying signal isolation attributes. Across the category-aligned scenarios, it reliably determines whether the scene contains a single emitter or overlapping activity, and whether the signal occupies a narrow, medium, or wide portion of the displayed band. This behavior provides an empirical evidence that RF structure learned from synthetic spectrograms can transfer to real SDR captures.

Furthermore, Table~\ref{tab:plr} presents PLR and hallucination counts, to showcase the degree of reliance on prompt metadata. RF-GPT's low PLR indicates that its bandwidth estimates are usually grounded in the waterfall image, rather than copied from the sample-rate field in the prompt. Importantly, a bandwidth estimate equal to the sample rate does not constitute leakage when the signal visibly occupies the full frequency span in the waterfall image, in which case the correct image-based bandwidth and the prompt sample rate are numerically identical. The leakage flag is applied only when the model reproduces the prompt value without image-based justification or explicitly attributes the bandwidth to the analyzer setting. On the other hand, the higher PLR of Llama~3.2~Vision indicates that it often relies on metadata when estimating bandwidth, especially when the signal is not clearly defined in the spectrum.

The untuned Qwen2.5-VL baseline exhibits a different failure mode, characterized by false-negative hallucinations. In many trials, it fails to detect a signal even when one is clearly visible in the waterfall. This indicates a mismatch between general visual pretraining and RF spectrogram semantics. A blue-to-yellow power map does not correspond to natural-image lighting, instead, it encodes received signal power. Without RF-specific adaptation, bright spectral regions are susceptible to misinterpretation as background texture, rather than representations of signal energy. Within this context, Llama~3.2~Vision produced fewer false negatives, but its successful cases are limited to scenarios when the signal is visually dominant, indicating that general-purpose visual features help only in high-contrast cases.

It is worth highlighting that, the obtained results showed that RF-GPT's main limitation is not physical geometry, but rather technology labeling. Several technology-label hallucinations follow a consistent pattern. At 98\,MHz, the model may assign labels such as DVB-S2 or 5G NR to a signal which spectral envelope matches flat-top waveform classes that are seen during training. At 806\,MHz and 950\,MHz, it may map cellular-like spectra to 5G NR even when the local band context makes LTE or GSM/LTE more reasonable. These errors show that the model has learned transferable spectral shapes, but it has not fully learned frequency-band constraints. Therefore, in real deployment, RF-GPT's protocol label should be interpreted as a hypothesis rather than a validated classification in the absence of frequency-allocation priors.

Finally, the killer test exposes a second limitation pertinent to low-SNR burst generalization. The RF-shielding bag attenuates the Bluetooth mouse transmission so that the expected narrow hopping bursts appear only faintly above the noise floor. RF-GPT detects intermittent activity, but over-segments the scene and misclassifies it as multiple overlapping signals. The untuned baselines experience a severe performance degradation by reporting no signal. This result indicates that the current synthetic training distribution covers open-air RF scenes more effectively than attenuated or shielded
emitters. Hence, adding low-SNR, narrow-band, and through-material examples to the synthetic dataset is a direct path to improving the model's robustness.

\subsection{Discussion and Limitations}
\label{sec:discussion}

The demonstrated results show that RF-GPT's synthetic training generalizes most reliably to physical signal geometry, with attributes such as occupied bandwidth, temporal activity, SNR regime, and signal isolation primarily derived from the real waterfall images rather than inferred from prompt metadata. Meanwhile, untuned VLMs frequently fail to detect visible RF energy or rely on textual measurement settings, highlighting the need for domain adaptation for reliable spectrogram grounding. The main remaining limitation is on semantic labeling, where RF-GPT can recognize the shape of a signal while assigning an invalid protocol name when frequency-band context is missing. The Bluetooth stress-test further identifies low-SNR burst activity as a boundary case, because attenuation removes the bright hopping patterns emphasized in the synthetic corpus. These observations suggest that future RF language models should combine synthetic waveform diversity with explicit band-allocation priors and low-SNR augmentation, while treating protocol labels as hypotheses unless independently verified.

\section{Conclusion}
\label{sec:conclusion}

This paper investigated whether an RF vision-language model trained only on synthetic spectrograms can support real over-the-air spectrum understanding. Using RF-Analyzer, we evaluated RF-GPT on live SDR captures and compared it with untuned vision-language baselines under identical input conditions. The results show that synthetic RF instruction tuning transfers to real data for physical signal interpretation, enabling reliable extraction of spectral structure and attributes. However, this transfer is not uniform. RF-GPT is more reliable for describing physical signal characteristics than for precise protocol identification, and performance degrades under low-SNR conditions. These results highlight the potential as well as the limitations of synthetic training, motivating future work on improved data coverage, band-aware conditioning, and validation mechanisms for robust RF spectrum analysis.


\bibliographystyle{IEEEtran}
\bibliography{references}

\end{document}

%% file: figures/architecture.tex
\begin{tikzpicture}[
  >=stealth, semithick,
  inner sep=4pt,
  hw/.style    ={draw=orange!70!black, fill=orange!10, rounded corners=3pt,
                 minimum height=0.85cm, align=center, font=\scriptsize},
  sim/.style   ={draw=gray!55,         fill=gray!12,   rounded corners=3pt,
                 minimum height=0.85cm, align=center, font=\scriptsize},
  acq/.style   ={draw=teal!70!black,   fill=teal!10,   rounded corners=3pt,
                 minimum height=0.85cm, align=center, font=\scriptsize},
  disp/.style  ={draw=blue!65,         fill=blue!8,    rounded corners=3pt,
                 minimum height=0.85cm, align=center, font=\scriptsize},
  ai/.style    ={draw=violet!65!black, fill=violet!10, rounded corners=3pt,
                 minimum height=0.85cm, align=center, font=\scriptsize},
  app/.style   ={draw=gray!55,         fill=gray!18,   rounded corners=3pt,
                 minimum height=0.85cm, align=center, font=\scriptsize},
  cont/.style  ={draw=gray!45, dashed, rounded corners=4pt, inner sep=5pt},
  arr/.style   ={->, semithick},
  darr/.style  ={->, semithick, dashed, gray!70},
  lbl/.style   ={font=\tiny, fill=white, inner sep=1pt},
]

\node[app, text width=8.0cm] (mainwin) at (4.3, 0.56) {%
  \textbf{MainWindow} $\cdot$ \textbf{PyQt6}\\%
  controls $\cdot$ chat interface $\cdot$ results display $\cdot$ frequency presets};

\node[ai, text width=3.7cm] (aimanager) at (2.0, 2.0) {%
  \textbf{AIClassifierManager}\\%
  \textbf{QThread}\\%
  image render $\cdot$ queue dispatch};

\node[ai, text width=1.55cm] (ollama) at (5.5, 2.0) {%
  \textbf{Ollama}\\Llama~3.2-V};

\node[ai, text width=1.55cm, right=0.2cm of ollama] (rfgpt) {%
  \textbf{RFGPT}\\Qwen2.5-VL};

\node[cont, fit=(ollama)(rfgpt), label={[font=\tiny,gray,yshift=2pt]above:AI backends (switchable)}] (bkcont) {};

\node[disp, text width=3.3cm] (spectrum) at (2.0, 3.6) {%
  \textbf{Spectrum plot}\\FFT $\cdot$ peak hold $\cdot$ avg};

\node[disp, text width=3.3cm] (waterfall) at (6.3, 3.6) {%
  \textbf{Waterfall}\\200-row time--freq heatmap};

\node[acq, text width=7.8cm] (receiver) at (4.3, 5.15) {%
  \textbf{USRPReceiver} $\cdot$ \textbf{QThread}\\%
  IQ capture $\cdot$ Hanning window $\cdot$ FFT (GPU/CPU) $\cdot$ power (dB)};

\node[cont, minimum width=8.4cm, minimum height=1.1cm] (hwcont) at (4.3, 6.8) {};

\node[hw, text width=2.6cm] (usrp) at (1.9, 6.8) {%
  \textbf{USRP B210}\\70\,MHz -- 6\,GHz};

\node[font=\tiny] at (4.3, 6.8) {\textit{or}};

\node[sim, text width=2.6cm] (simusrp) at (6.7, 6.8) {%
  \textbf{Simulated}\\software sim};


\draw[arr] (hwcont.south) -- (receiver.north)
  node[lbl, right, midway, xshift=2pt] {\tiny IQ samples};

\draw[arr] ([xshift=-2.3cm]receiver.south) -- ++(0,-0.28)
  -| (spectrum.north);

\draw[arr] ([xshift=2.0cm]receiver.south) -- ++(0,-0.28)
  -| (waterfall.north);

\draw[arr] (waterfall.south) -- ++(0,-0.35) -| (aimanager.north)
  node[lbl, above, pos=0.6] {waterfall buffer};

\draw[arr] (aimanager.east) -- (bkcont.west)
  node[lbl, above, midway] {PNG};

\draw[arr] (ollama.south) -- (ollama.south |- mainwin.north);

\draw[arr] (rfgpt.south) -- (rfgpt.south |- mainwin.north);

\draw[darr] ([xshift=-3.0cm]mainwin.north)
  -- ([xshift=-3.0cm]mainwin.north |- aimanager.south)
  node[lbl, right, near end] {\color{gray}\tiny user trigger};

\begin{scope}[shift={(0.0, -0.6)}] 
  \node[hw,  minimum width=0.35cm, minimum height=0.2cm, inner sep=0] (l1) at (0.0, 0) {};
  \node[font=\tiny, anchor=west] at (0.22, 0) {Hardware};

  \node[acq, minimum width=0.35cm, minimum height=0.2cm, inner sep=0] at (1.6, 0) {};
  \node[font=\tiny, anchor=west] at (1.82, 0) {Acquisition / DSP};

  \node[disp,minimum width=0.35cm, minimum height=0.2cm, inner sep=0] at (3.8, 0) {};
  \node[font=\tiny, anchor=west] at (4.02, 0) {Display};

  \node[ai,  minimum width=0.35cm, minimum height=0.2cm, inner sep=0] at (5.3, 0) {};
  \node[font=\tiny, anchor=west] at (5.52, 0) {AI engine};

  \node[app, minimum width=0.35cm, minimum height=0.2cm, inner sep=0] at (6.8, 0) {};
  \node[font=\tiny, anchor=west] at (7.02, 0) {Application shell};
\end{scope}

\end{tikzpicture}

%% file: references.bib
@article{swami2000hierarchical,
  author    = {Swami, Ananthram and Sadler, Brian M},
  title     = {Hierarchical digital modulation classification using cumulants},
  journal   = {IEEE Transactions on Communications},
  volume    = {48},
  number    = {3},
  pages     = {416--429},
  year      = {2000}
}

@article{oshea2018over,
  author    = {O'Shea, Timothy James and Roy, Tamoghna and Clancy, T Charles},
  title     = {Over-the-air deep learning based radio signal classification},
  journal   = {IEEE Journal of Selected Topics in Signal Processing},
  volume    = {12},
  number    = {1},
  pages     = {168--179},
  year      = {2018}
}

@inproceedings{west2017deep,
  author       = {West, Nathan E and O'Shea, Tim},
  title        = {Deep neural network architectures for modulation classification},
  booktitle    = {2017 IEEE 18th Wireless and Microwave Technology Conference (WAMICON)},
  pages        = {1--6},
  year         = {2017}
}

@article{zhou2025llmtelecom,
  author  = {Zhou, H and others},
  title   = {Large Language Model ({LLM}) for Telecommunications: A Comprehensive Survey on Principles, Key Techniques, and Opportunities},
  journal = {IEEE Communications Surveys \& Tutorials},
  volume  = {27},
  number  = {3},
  year    = {2025}
}

@article{zou2025seeingradiokhalifa,
  author  = {Zou, Hang and others},
  title   = {Seeing Radio: From Zero {RF} Priors to Explainable Modulation Recognition with Vision Language Models},
  journal = {arXiv preprint arXiv:2601.13157},
  year    = {2026}
}

@article{zou2025telecomgpt,
  author  = {Zou, Hang and others},
  title   = {{TelecomGPT}: A Framework to Build Telecom-Specific Large Language Models},
  journal = {IEEE Transactions on Machine Learning in Communications and Networking},
  volume  = {3},
  pages   = {948--975},
  year    = {2025}
}

@article{bariah2024largegenerativeai,
  author  = {Bariah, Lina and others},
  title   = {Large Generative {AI} Models for Telecom: The Next Big Thing?},
  journal = {IEEE Communications Magazine},
  volume  = {62},
  number  = {11},
  year    = {2024}
}

@article{boegner2022largescalerf,
  author  = {Boegner, Luke and others},
  title   = {Large scale radio frequency signal classification},
  journal = {arXiv preprint arXiv:2207.09918},
  year    = {2022}
}

@misc{rohdeschwarz2024,
  author       = {{Rohde \& Schwarz}},
  title        = {Spectrum Analyzers and Signal Analyzers},
  howpublished = {\url{https://www.rohde-schwarz.com/us/products/test-and-measurement/benchtop-analyzers/rs-fsc-spectrum-analyzer_63493-10891.html}},
  year         = {2024},
  note         = {Accessed: May 2025}
}

@misc{keysight2024,
  author       = {{Keysight Technologies}},
  title        = {Signal Analyzers},
  howpublished = {\url{https://www.keysight.com/us/en/product/N9000B/cxa-signal-analyzer-multi-touch-9-khz-26-5-ghz.html}},
  year         = {2024},
  note         = {Accessed: May 2025}
}

@article{zou2026rf,
  title={{RF-GPT: Teaching AI to See the Wireless World}},
  author={Zou, Hang and Tian, Yu and Wang, Bohao and Bariah, Lina and Lasaulce, Samson and Huang, Chongwen and Debbah, M{\'e}rouane},
  journal={arXiv preprint arXiv:2602.14833},
  year={2026}
}

@inproceedings{kwon2023efficient,
  title={{Efficient memory management for large language model serving with PagedAttention}},
  author={Kwon, Woosuk and others},
  booktitle={Proceedings of the 29th symposium on operating systems principles},
  pages={611--626},
  year={2023}
}
